\documentclass[journal,twoside,web]{ieeecolor}
\usepackage{jsen}
\usepackage{cite}
\usepackage{amsmath,amssymb,amsfonts}
\usepackage{algorithmic}
\usepackage{graphicx}
\usepackage{textcomp}
\usepackage{wrapfig}
\usepackage{multirow}
\usepackage{url}
\usepackage{bm}	

\newcommand{\update}[1]{{#1}}
\newcommand{\delete}[1]{}

\newcommand{\argmax}{\mathop{\rm argmax}\limits}
\newcommand{\argmin}{\mathop{\rm argmin}\limits}
\newcommand{\ub}[1]{\underline{\textbf{#1}}}

\newcommand{\um}[1]{\textbf{#1}}

\def\BibTeX{{\rm B\kern-.05em{\sc i\kern-.025em b}\kern-.08em
    T\kern-.1667em\lower.7ex\hbox{E}\kern-.125emX}}
\definecolor{abstractbg}{rgb}{0.89804,0.94510,0.83137}
\begin{document}
© 2020 IEEE.  Personal use of this material is permitted.  Permission from IEEE must be obtained for all other uses, in any current or future media, including reprinting/republishing this material for advertising or promotional purposes, creating new collective works, for resale or redistribution to servers or lists, or reuse of any copyrighted component of this work in other works.
\newpage

\title{Smartphone Sensor-based Human Activity Recognition Robust to Different Sampling Rates}

\author{Tatsuhito Hasegawa, \IEEEmembership{Member, IEEE}
\thanks{Manuscript received September 18, 2020; replied the first decision October 11; received again October 20, 2020; date of current version October 24, 2020. This work was supported by the JSPS KAKENHI under Grant 19K20420 and Tateisi Science and Technology Foundation research grant (A). A basic experiment of this work is conducted by Mr. Hirofumi Kimura. (Corresponding author: Tatsuhito Hasegawa). }
\thanks{Tatsuhito Hasegawa is with Graduate School of Engineering, University of Fukui, Fukui 910-8507, Japan. (e-mail: t-hase@u-fukui.ac.jp). }
\thanks{Digital Object Identifier 10.1109/JSEN.2020.3038281}
}

\IEEEtitleabstractindextext{%
\fcolorbox{abstractbg}{abstractbg}{%
\begin{minipage}{\textwidth}%
\begin{wrapfigure}[12]{r}{3in}%
\includegraphics[width=2.8in]{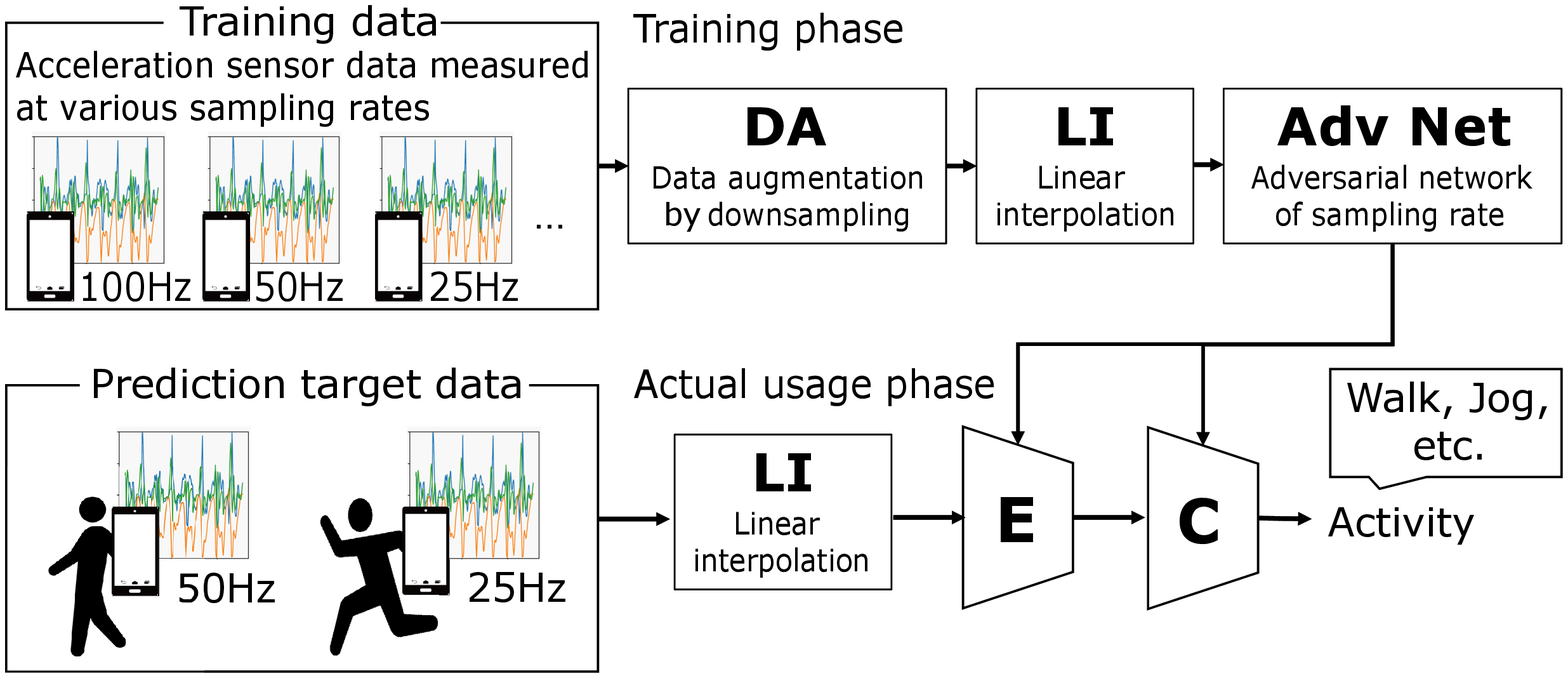}%
\end{wrapfigure}%
\begin{abstract}
There is a research field of human activity recognition that automatically recognizes a user's physical activity through sensing technology incorporated in smartphones and other devices. When sensing daily activity, various measurement conditions, such as device type, possession method, wearing method, and measurement application, are often different depending on the user and the date of the measurement. Models that predict activity from sensor values are often implemented by machine learning and are trained using a large amount of activity-labeled sensor data measured from many users who provide labeled sensor data. However, collecting activity-labeled sensor data using each user's individual smartphones causes data being measured in inconsistent environments that may degrade the estimation accuracy of machine learning. In this study, I propose an activity recognition method that is robust to different sampling rates---even in the measurement environment. The proposed method applies an adversarial network and data augmentation by downsampling to a common activity recognition model to achieve the acquisition of feature representations that make the sampling rate unspecifiable. Using the Human Activity Sensing Consortium (HASC), which is a dataset of basic activity recognition using smartphone sensors, I conducted an evaluation experiment to simulate an environment in which various sampling rates were measured. As a result, I found that estimation accuracy was reduced by the conventional method in the above environment and could be improved by my proposed method.
\end{abstract}

\end{minipage}}}

\maketitle

\section{Introduction}
\label{sec:introduction}
\IEEEPARstart{W}{ith} the widespread use of smartphones and IoT devices, many context-awareness services, which allow computers to recognize human activity and the surrounding environment, have been studied. Among them, several studies have been conducted on human activity recognition, which uses sensors on smartphones and other devices to automatically recognize a user's physical activity \cite{Kwapisz2011, Xu2015, Shoaib2016, Robert2019}. Activity recognition can be applied to various fields, such as lifelogs; automatically change a smartphone's settings according to a user's behavior; and marketing and conducting an analysis of group behavior by collecting large-scale user data.

Although there are image-based methods for recognizing human activity that use stationary cameras \update{\cite{Liu2019}}, I focus on human activity recognition using sensor data measured by the devices users carry with them on a daily basis. Compared to stationary-camera methods, sensor-based methods have the advantages of being able to acquire data at any time and place as well as acquire information specific to each user. The disadvantage is that each user must own the sensor device, but this issue can be avoided by using smartphones and smartwatches, the ownership of which has rapidly spread in recent years. However, there remains the problem of various measurement conditions, such as device type, possession method, wearing method, and measurement application, which vary from user-to-user and from one measurement date to the next.

As Stisen et al. \cite{Allan2015} mentioned, in the case of activity recognition using smartphones, sensor characteristics, such as sampling rate, vary between devices. In addition, the API\footnote{Android developers SensorManager: \url{https://developer.android.com/reference/android/hardware/SensorManager}} of Android smartphones allows us to change the sampling interval of sensors. Therefore, in activity recognition using smartphones, there is a possibility that various sampling rates may be included in both training data and prediction-target data when the sensor data is measured in an actual environment by many users.

The automatic classification of sensor data by machine learning is a common activity recognition principle \cite{Avci2010, Chen2012, Lala2012, wang2019}. A unified measurement environment is desirable, because training a model with inconsistent data may reduce the estimation accuracy of machine learning. Although deep learning requires a large amount of data for training, it is not easy to collect such data in a unified measurement environment. Therefore, a method that can achieve high recognition accuracy, even for inconsistent data, is desired.

This study aims to develop an activity recognition method that is robust to different sampling rates. I propose an adversarial network of sampling rates and data augmentation (DA) via downsampling to improve the accuracy of activity recognition in an environment in which sensors measure data at various sampling rates; this will hereafter be referred to as an ``SR mixed environment.''

\section{Related works}
\subsection{Sampling rates in conventional activity recognition studies}
\delete{
Sampling rates are not discussed in many of the activity recognition studies referenced in Section I \cite{Kwapisz2011, Robert2019, Shoaib2016, Avci2010, Chen2012}. The two studies that described the sampling rate also used data that was uniformly measured at a constant sampling rate \cite{Lala2012, Xu2015}. Li et al. \cite{Li2018} compared and verified multiple representation-learning-based activity recognition methods using several benchmark datasets of sensor-based activity recognition; they used the OPPORTUNITY \cite{Ricardo2013} and UniMiB-SHAR (University of Milano Bicocca Smartphone-based HAR) \cite{Daniela2017} datasets, which were measured at a unified sampling rate---30 Hz and 50 Hz, respectively.
}
\update{
I conducted a comprehensive survey of studies on activity recognition using sensors. Many studies \cite{Kourosh1998, Aminian1999, Bao2004, Ravi2005, Shoaib2016, Robert2019} have used machine learning techniques (e.g., decision tree) to recognize activities of daily living (ADL; e.g., sitting and walking) by calculating feature vectors (e.g., mean and variance) extracted from raw sensor data. Additionally, Xu et al. \cite{Xu2015} proposed gesture recognition with a similar method. These studies have been conducted to construct an original sensor dataset. Although one study \cite{Kourosh1998} did not specify a sampling rate, most studies adopted a uniform sampling rate within the study (\cite{Aminian1999} 10Hz, \cite{Bao2004} 76Hz, \cite{Ravi2005} 50Hz, \cite{Shoaib2016} 50Hz, \cite{Robert2019} 20Hz and \cite{Xu2015} 128Hz).
}

\update{
Many studies have constructed and released large-scale benchmark datasets to the public. The OPPORTUNITY dataset \cite{Roggen2010, Ricardo2013} collected sensor data over a long period of time by having people wear a large number of sensors. It was used for multiple tasks, such as ADL recognition, segmentation, and gestures recognition. It is provided as a benchmark dataset. The sampling rate of OPPORTUNITY was limited to 64 Hz or 32 Hz based on the measurement stage. There are many other benchmark datasets for ADL recognition, such as WISDM \cite{Kwapisz2011} (sampling rate is 20 Hz), Skoda \cite{Zappi2012} (100 Hz), PAMAP \cite{Reiss2012a} (100 Hz), Smartphone Dataset \cite{Anguita2013} (50 Hz), UniMiB SHAR \cite{Daniela2017} (50 Hz), mHealth \cite{Banos2014, Banos2015} (50 Hz), and Hand Gesture \cite{Bulling2014} (32 Hz). All of these datasets have uniform sampling rates, and their recognition methods involve simply extracting feature vectors and using machine learning.
}

The benchmark HASC dataset \cite{HASC} includes a large amount of sensor data measured in an SR mixed environment. The measurements are mainly made with smartphones, such as iPhone 3Gs, but also with a wide range of other devices, such as the sensor device WAA-001 \footnote{ATR-Promotions Inc. WAA-001: \url{https://www.atr-p.com/support/support-sensor01.html}}. The sampling rate was recorded under various measurement conditions, which ranged between 25-100 Hz. Wang's survey \cite{wang2019}, introduced two datasets (i.e., ActiveMiles \cite{Ravi2016} and Heterogeneous \cite{Allan2015}) that were measured in environments with different sampling rates. Ravi et al.'s study \cite{Ravi2016} collected data in an SR mixed environment and published them as ActiveMiles datasets; however, they did not implement any innovations to improve the effect of the SR mixed environment. \update{Heterogeneous is described below.}

\update{
\subsection{Activity recognition by deep learning}
There are many studies that use deep learning for sensor-based activity recognition. Some studies have conducted their own measurement experiments; however, all of them use a uniform sampling rate (\cite{Chen2015} 100 Hz, \cite{Zhou2019} 50 Hz, \cite{Gumaei2019} 50 Hz). As deep learning models, a model \cite{Chen2015, Zhou2019} that alternates convolution and pooling layers in a simple manner and a model \cite{Gumaei2019} that uses simple recurrent units and gated recurrent units (GRU) together have been proposed.
}

\update{
Some studies have proposed deep learning models and evaluated them using benchmark datasets. Yang et al. \cite{Yang2015} evaluated a simple convolutional neural network (CNN) using the OPPORTUNITY and Hand Gesture datasets. Ha et al. \cite{Ha2015} evaluated 1D and 2D convolution using mHealth and Skoda datasets. Ord\'{o}\~{n}ez et al. \cite{Francisco2016} proposed a model that employed LSTM after a multistage convolution and evaluated it using the OPPORTUNITY, PAMAP, Skoda, and mHealth datasets. Li et al. \cite{Li2018} also evaluated a model that employed LSTM after convolution and pooling using the OPPORTUNITY and UniMiB SHAR datasets. Yang et al. \cite{Yang2018} proposed a method that used multiple feature extractor CNN for each inertial measurement unit (IMU) and weighted each feature map and evaluated it using the OPPORTUNITY, PAMAP and UniMiB SHAR datasets. Xu et al. \cite{Xu2019} proposed a method to use human-designed features (hand-crafted features, or HCF) together with GRUs to connect to the Inception modules and evaluated it using the OPPORTUNITY, PAMAP and Smartphone datasets. These studies were conducted using datasets with uniform sampling rates, and no effort was made to address variable sampling rates.} In my previous study \cite{hasegawa2019}, I conducted a comparative verification of multiple \update{CNN} models using the HASC dataset, but I limited the evaluation to data with a sampling rate of 100 Hz.

\update{
\subsection{Activity recognition with different sampling rate}
}
Allan et al. \cite{Allan2015} focused on differences in sampling rates and created the Heterogeneous dataset. They experimentally showed that the sensor characteristics differed from device-to-device, and they published a dataset of the everyday activities collected from the sensors of the various devices. They also described the adverse effects of an SR mixed environment and proposed a method to train classifiers for each cluster after the clustering of sensor characteristics. However, their method required that the sensor characteristics of many devices be measured in advance. \update{Although Ma et al. \cite{Ma2019} evaluated the Heterogeneous dataset, they proposed a model structure for deep learning and did not focus on variable sampling rates in method development.}

Nakajima et al. \cite{nakashima2011} proposed a method to lessen the power consumption of activity recognition by deliberately controlling the sampling rate. Their method achieved power saving without diminishing the activity-recognition accuracy by limiting the sampling rate and interpolating the sensor data. Although this is similar to my study in that it interpolates low-sampling-rate data, their study differs from mine because it does not target an SR mixed environment. In addition, my proposed method includes a process to align the sampling rate by upsampling, which can then be replaced by the Nakajima et al.'s method. In other words, my proposed method can coexist with various interpolation methods. \update{Additionally, Qi et al. \cite{Qi2013} proposed a method to reduce the overall sampling rate by suitably switching to use two classifiers for low and high sampling rates. Khan et al. \cite{Khan2016} also proposed a sampling rate optimization method from unlabeled data using statistical tests.}

\subsection{Adversarial network}
In this study, I propose an activity recognition model that is robust to differet sampling rates by applying an adversarial deep learning network. Here, I describe the relevant studies on the adversarial network.

The Generative Adversarial Network (GAN) \cite{goodfellow2014}, which was proposed by Goodfellow et al., is a representative study using adversarial network. The GAN is composed of two models---a generator that generates realistic images and a discriminator that classifies the images as real or automatically generated---and the two models are adversarially trained. 

Domain-Adversarial Training of Neural Networks (DANN) \cite{DANN2016} and Adversarial Discriminative Domain Adaptation (ADDA) \cite{ADDA2017} were proposed by studies that applied an adversarial network in the context of domain adaptation. These methods target a domain-adaptation problem that uses a source domain dataset with labels to estimate an unlabeled target domain dataset. The basic principles are to train a network that maps the data in the source and target domains to the latent space and to antagonize the Discriminator, which distinguishes the domain from the feature representation derived at that time.

An example of an application of the adversarial network in the field of activity recognition is the study by Iwasawa et al. \cite{iwasawa2017}, which highlighted the possibility that activity recognition could include privacy information that identifies the user. Their model of representation learning removes individually identifiable features from sensor data without reducing the accuracy of the activity recognition. This is not done to improve the accuracy of the activity recognition but rather to acquire feature representations that do not identify the individual user. \update{Bai et al. also proposed an adversarial training method for activity recognition and individual discrimination \cite{Bai2020}. They experimentally showed that adversarial training improves activity-recognition accuracy and proposed the use of a multi-view convolutional method.}

\subsection{Contributions of study}
Based on the above-cited related works, the four main contributions of this study are as follows:

\begin{itemize}
\item \update{{\bf Clarify the negative}} {\bf effect of an SR mixed environment.} Most of the previous studies used data measured at a uniform sampling rate. In this study, I experimentally clarify the \update{negative} effect of the SR mixed environment on the accuracy of the activity recognition.
\item \update{{\bf Propose}} {\bf an adversarial network} \update{{\bf for}} {\bf an SR mixed environment} \update{{\bf and determine the effective scenarios}}. This model is composed of an activity classifier and a sampling-rate discriminator. I also formulate and implement this model. As an experimental result, the adversarial network contributes to improving the accuracy in three scenarios: the prediction-target data is measured at high sampling rates; the training data includes more low sampling rate data; and the number of subjects in the training data is low.
\item \update{{\bf Propose a}} {\bf DA method by downsampling} \update{{\bf for an SR mixed environment and determine the effective scenarios}}. The data measured in an SR mixed environment is augmented to various sampling rates via downsampling. Although activity recognition generally involves pre-processing to unify sampling rates, the novelty of this method is that the data is intentionally augmented to different sampling rates via downsampling. As an experimental result, DA contributes to improving the accuracy in two scenarios: the prediction-target data is measured at low sampling rates and the training data includes more high sampling rate data.
\end{itemize}

\section{Activity recognition robust to different sampling rates}
\subsection{Overview}
\begin{figure}[tb]
\begin{center}
\includegraphics[width=8.5cm]{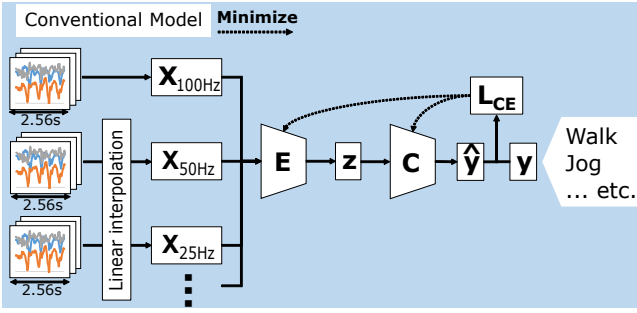}
\caption{Conventional method: common method of human activity recognition using CNN. }
\label{fig:conventional}
\end{center}
\end{figure}

My proposed method varies from a conventional method in the following three ways:
\begin{itemize}
\item Uses both activity labels and sampling rate labels
\item Uses the sampling rate labels to adversarially train the model
\item Uses the DA by downsampling
\end{itemize}
Deep-learning-based activity recognition models that were adopted in previous studies, such as the work of Li et al. \cite{Li2018}, are models that simply input sensor data through a network and classify activities. There are various networks, such as Multi-Layer Perceptron (MLP), Convolutional Neural Network (CNN), and Recurrent Neural Network (RNN). In this study, I discuss CNN-based networks. As described above, the models in related studies are trained under limited to the unified sampling rate. In order to be applied in an SR mixed environment, I would therefore need to train the model using various sampling rates, as illustrated in Fig. \ref{fig:conventional}, where $\bm{X}_{\rm 100Hz}$ is the sensor data measured at a sampling rate of 100 Hz, and is then an input in the network; $E$ is a feature extractor that is generally composed of CNN and accepts sensor data measured at various sampling rates; $\bm{z}$ is the feature map which is automatically extracted by the feature extractor from raw sensor data; $C$ is a classifier that classifies activities; and $\hat{\bm{y}}$ is the predicted result of the classifier. The loss function of the network is cross-entropy loss $L_{CE}$, which is calculated from the predicted result $\hat{\bm{y}}$ and the true activity label $\bm{y}$. Note that, to input the same time-length sensor data to the encoder $E$, the raw data needs to undergo linear interpolation to unify the size of the samples.

\begin{figure}[tb]
\begin{center}
\includegraphics[width=8.5cm]{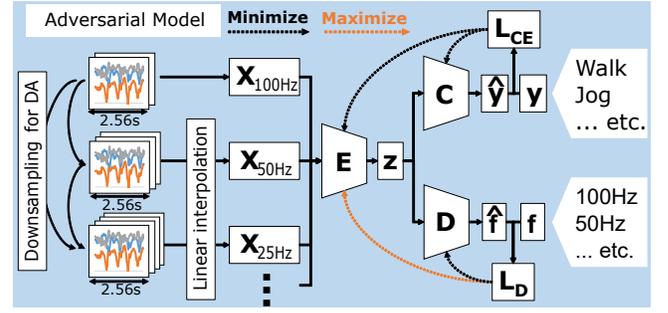}
\caption{Proposed method: human activity recognition that is robust to the different sampling rates. }
\label{fig:adversarial}
\end{center}
\end{figure}

An overview of my proposed method is shown in Fig. \ref{fig:adversarial}. First, my proposed method conducts DA through generating low-sampling rate sensor data from high-sampling rate sensor data via downsampling (details will be described in Section \ref{sec:da}). Next, the upper portion of Fig. \ref{fig:adversarial} indicates that the activity-recognition classifier is the same as the previous method; $E$ and $C$ are trained by minimizing loss function $L_{CE}$. The lower portion of Fig. \ref{fig:adversarial} indicates the discriminator part $D$, which estimates the sampling rate $\hat{\bm{f}}$ of the input sensor data from the latent vector $\bm{z}$; this is similar to the domain-detection component of the DANN domain adaptation method \cite{DANN2016}. In this model, the input length is unified via linear interpolation, but there is a difference in waveform smoothness between $X_{100Hz}$ and $X_{25Hz}$ with linear interpolation; $D$ estimates the sampling rate by detecting this difference. Discriminator error $L_{D}$ is calculated from the estimated result $\hat{\bm{f}}$ and the true sampling rate $\bm{f}$ as a loss function of the network. 

The advantages of my proposed method are that the feature representation can be robust to different sampling rates through using an adversarial network and low-sampling rate sensor data augmented by downsampling.

\update{
There are various methods to acquire universal feature representation. Maximum mean discrepancy (MMD) \cite{Gretton2012} is a method for calculating the difference between two probability distributions using the kernel method, and there is a method to acquire a universal feature representation by minimizing MMD between two domain datasets \cite{Tzeng2014, Long2015}. There is also a method that minimizes the distance between the mean and variance of two domains in batch normalization layer \cite{AdaBN2018}. Although these methods may work well to acquire a universal feature representation for different sampling rates, I employed adversarial training using discriminator in this study because of its ease of imlementation.}

\subsection{Formulation}
$L_{CE}$ is a common loss function for classification tasks used in conventional neural networks. It is defined by the following formula:
\begin{equation}
\label{eq:lce}
L_{CE} = -\frac{1}{N}\sum_{n=1}^{N}\sum_{m=1}^{M}\{y_{nm} \log{C(E(x_n))} \}
\end{equation}
where $N$ and $M$ denote the size of the minibatch and the number of categories  of the activity labels, respectively, $x_n$ denotes the $n$-th input series of the minibatch of input $X_{\rm 100Hz}$, and $y_{nm}$ denotes the $m$-th value of a one-hot vector of the $n$-th output. In the conventional method, we must prepare $C$ and $E$ for each sampling rate (e.g., $C_{\rm 100Hz}$).

$L_D$ is a discriminator loss function that is defined by the following formula:
\begin{equation}
\label{eq:ld}
L_D = -\frac{1}{N}\sum_{n=1}^{N}\sum_{k=1}^{K}\{f_{nk} \log{D(E(x_n))} \}
\end{equation}
where $K$ denotes a number of the kind of sampling rates, and $L_D$ signifies the cross-entropy loss function of discriminator $D$ that classifies the sampling rate. DANN \cite{DANN2016} adopted binary cross-entropy as a loss function to distinguish between the source and target domains. On the other hand, assuming that the measured sensor data in this study includes various sampling rates, I define this problem as a classification of sampling rates. Therefore, I use cross-entropy loss as a discriminator loss function. 

I finally define loss function $L$ for the entire network of my proposed method as the following formula:
\begin{equation}
\label{eq:l}
L = L_{CE} - \lambda L_D
\end{equation}
Iwasawa et al. \cite{iwasawa2017} pointed out that using categorical cross-entropy as a discriminator loss $L_D$ reduces the recognition ability of $D$ more than the use of binary cross-entropy, thereby making it more difficult to balance the adversity of $C$ and $D$. Thus, they applied annealing to gradually increase $\lambda$. Since this phenomenon did not occur in a preliminary experiment of my study, I set $\lambda = 1$.

\subsection{Optimization}
My proposed method consists of three models: a feature extractor $E$, an activity classifier $C$, and a sampling-rate discriminator $D$. Assuming the parameters of each network are $\theta_E$, $\theta_C$, and $\theta_D$, each parameter for $\hat{\theta_E}$, $\hat{\theta_C}$, and $\hat{\theta_D}$ to be searched is calculated by the following formulas:
\begin{equation}
\label{eq:opt1}
(\hat{\theta_E}, \hat{\theta_C}) = \argmin_{\theta_E, \theta_C} L(\theta_E, \theta_C, \hat{\theta_D})
\end{equation}
\begin{equation}
\label{eq:opt2}
\hat{\theta_D} = \argmax_{\theta_D} L(\hat{\theta_E}, \hat{\theta_C}, \theta_D)
\end{equation}

In this study, the above minimization and maximization problems are implemented via alternating optimization. In Eq. (\ref{eq:opt1}), better parameters $\theta_E$ and $\theta_C$ are searched by minimizing $L$ computed with fixed $\theta_D$. This process searches such parameters that minimize $L_{CE}$ and maximize $L_D$, and it is equivalent to training $E$ and $C$ to improve the accuracy of the activity recognition and training $E$ to decrease the discrimination accuracy of the sampling rate. In Eq. (\ref{eq:opt2}), better parameters $\theta_D$ are searched by maximizing $L$ computed with fixed $\theta_E$ and $\theta_C$. At this time, since $L_{CE}$ is constant due to parameters $\theta_E$ and $\theta_C$ being fixed, the maximization of $L$ is equivalent to the minimization of $L_D$. Thus, it is equivalent to training $D$ to improve the discrimination accuracy of the sampling rate. These parameters are alternately optimized in minibatch units.
\begin{figure}[b]
\begin{center}
\includegraphics[width=9cm]{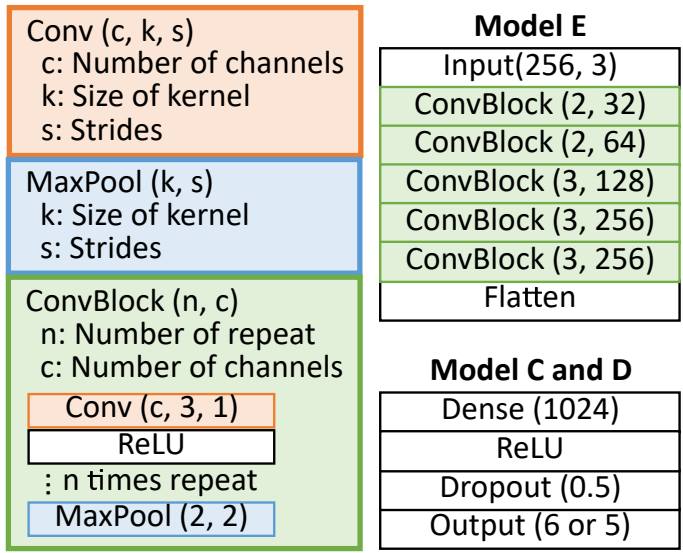}
\caption{Model architecture based on VGG16, which I used in experiments. }
\label{fig:vgg}
\end{center}
\end{figure}

\subsection{Architecture for each model}
My proposed method consists of three models: $E$, $C$, and $D$; the architecture of each model is outside the scope of the present study. In this study, I adopt a model based on VGG16, which is an easy-to-implement architecture that achieves high estimation accuracy for activity recognition. I design $E$ up to the Flatten layer of VGG16, and $C$ and $D$ are simply applied to fully connected layers (i.e., 1024 units) with activation function ReLU and dropout 0.5 (see Fig. \ref{fig:vgg}). The output layer of $C$ has six units (i.e., six categories of activity), and $D$ has five units (i.e., five sampling-rate categories).

\subsection{Data augmentation by downsampling}
\label{sec:da}
In general, most activity-recognition studies have targeted sensor data measured at a unified sampling rate. In contrast, my proposed method is to intentionally downsample the sensor data measured in an SR mixed environment for DA toward a more-mixed sampling rate. I assume that the DA improves the estimation accuracy for the sensor data measured at a low sampling rate. In addition, the augmentation of the data measured at various sampling rates makes it possible for the model to acquire a robust representation for different sampling rates.

When $X_{100Hz}$, $X'_{50Hz}$, and $X''_{25Hz}$ data are given as training data (Fig. \ref{fig:adversarial}), downsampling can simulate a dataset with a lower sampling rate than the dataset itself. For example, $X_{100Hz}$ can reproduce $X_{50Hz}$, $X_{25Hz}$, etc. via downsampling. By implementing the whole pattern, it is possible to augment the data, especially at low sampling rates.

In this study, a low-pass filter (LPF) is applied prior to the downsampling and uses thinning to prevent aliasing errors. Based on the sampling theorem, time-series waveform data should be sampled at a frequency greater than twice the highest frequency included in the signal to be measured. Therefore, when I downsample sensor data from 100 Hz to 50 Hz, the information contained in the signal should be less than 50 Hz / 2 = 25 Hz; to achieve this, I use an LPF to remove high-frequency components from the signal. 

\section{Experimental settings}
\subsection{Dataset}
I use the HASC dataset \cite{HASC}, which is an activity-recognition dataset provided by a non-profit voluntary organization for the purpose of constructing a large-scale database with wearable sensors in order to understand human activity. As a corpus, sensor data, including acceleration and gyro, labeled with six basic activities (i.e., stay, walk, jog, skip, up stairs (stUp), and down stairs (stDown)) is provided.

I extracted data at a sampling rate of 100 Hz for the BasicActivity corpus from the period of 2011--2013 and only use the raw sensor data of the accelerometer. For pre-processing, five seconds are trimmed from the start and end of each measurement file to remove the influence of the storing behavior of the device, and a time-series division is performed with 256 samples of frame size and 256 samples of stride. Meta information, such as device type and gender is not used. Data from 176 individuals for whom more than one frame of data could be obtained after trimming were used in my experiments.

Although the HASC includes data measured at sampling rates other than 100 Hz, the amount of the data is relatively small. In this study, in order to verify the effectiveness of the proposed method under various conditions, I reproduce an environment with mixed sampling rates from 100 Hz data.
\begin{figure*}[t]
\begin{center}
\includegraphics[width=17cm]{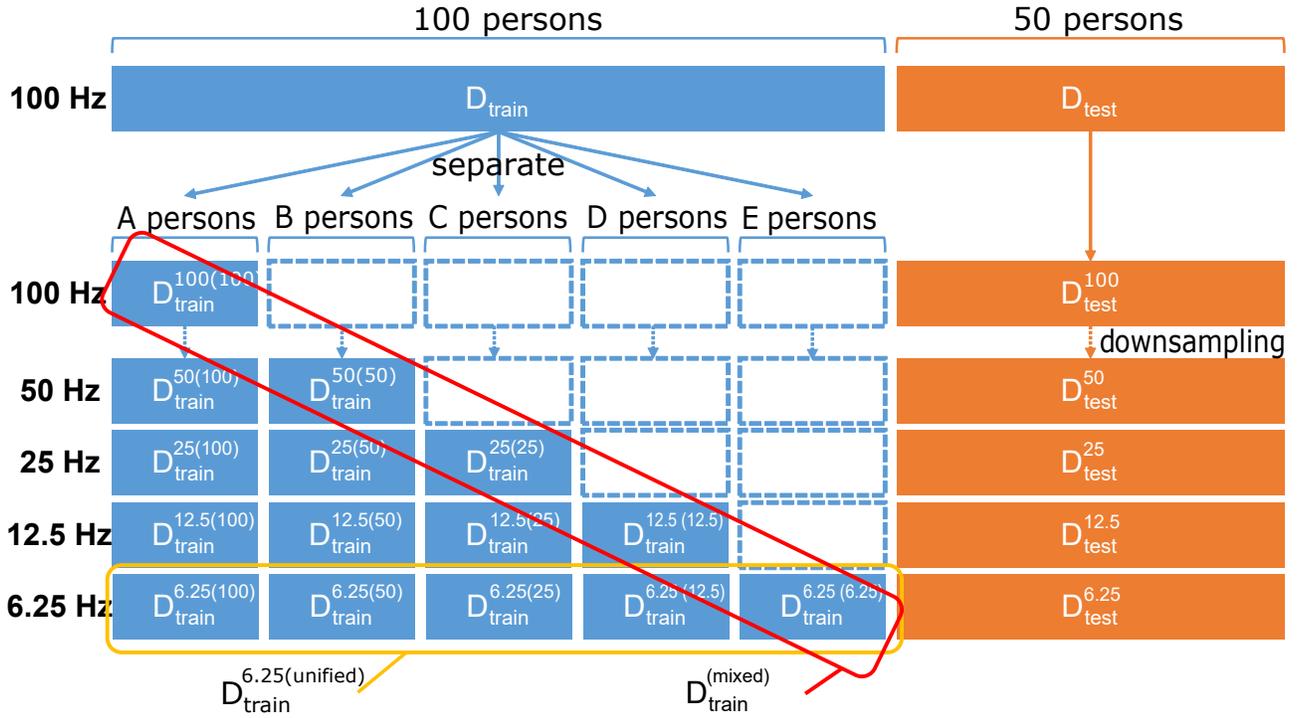}
\caption{How to separate the data for preparing training and testing dataset in SR mixed environment. }
\label{fig:dataset}
\end{center}
\end{figure*}

\subsection{Division of dataset}
\label{sec:env}
I describe the separation method of the dataset used for training and testing the model (Fig. \ref{fig:dataset}). In activity-recognition studies, it is known that the use of the test-users' individual data during training increases the estimation accuracy. The training and validation datasets in this study are therefore divided by the holdout method by person; an overview of the division is illustrated in Fig. \ref{fig:dataset}. From the dataset of 176 participants described in the previous section, 100 participants are randomly selected as the training dataset $D_{train}$; similarly, 50 participants are separately selected at random as the test dataset $D_{test}$.

Considering the various sampling rates of $D_{test}$, I prepare several types of $D_{test}$, ranging from 100 Hz ($D_{test}^{100}$) to 50 Hz ($D_{test}^{50}$), 25 Hz, 12.5 Hz, and 6.25 Hz via downsampling, and I verify the accuracy of each. I use 50 participants who are identical across sampling rates. The specific downsampling procedure is described in the next section. Note that this process is to reproduce an SR mixed environment but not augment the data.

The training dataset simulates an SR mixed environment by downsampling from $D_{train}$. First, I divide $D_{train}$ by a specific number of subjects (names A, B, C, D, and E in the Fig. \ref{fig:dataset}). Next, the data of A is assumed to be measured at a sampling rate of 100 Hz, and hence, it is called $D_{train}^{100(100)}$; and the data of B is assumed to be measured at a sampling rate of 50 Hz, and it is named $D_{train}^{50(50)}$. At this time, the 100 Hz data of B is included in $D_{train}$ itself, but it is only used to convert the data to $D_{train}^{50(50)}$ and not in the experiment in the 100 Hz state. Similarly, the data of C, D, and E are measured at 25 Hz, 12.5 Hz, and 6.25 Hz and called $D_{train}^{25(25)}$, $D_{train}^{12.5(12.5)}$, and $D_{train}^{6.25(6.25)}$, respectively. These data are collectively referred to as training dataset $D_{train}^{(mixed)}$ in the SR mixed environment. 

\subsection{Models and optimizers}
The two models used for comparison are the conventional CNN model (i.e., $M_{org}$) shown in Fig. \ref{fig:conventional}, and the adversarial network model with the DA of my proposed method (i.e., $M_{DA-Adv}$) shown in Fig. \ref{fig:adversarial}. The architecture of $M_{org}$ is equivalent to that of $M_{DA-Adv}$, where $E$ and $C$ are identical, and only $D$ does not exist. In this study, the optimization is carried out using Adam \cite{Kingma2015} with a learning rate of 0.0001, $\beta_1=0.9$, and $\beta_2=0.999$ for all models. Training is performed at 150 epochs. The estimation accuracy of the validation data $D_{test}$ was measured and recorded at every 10 epochs during the training.
\\
\par

\section{Experimental results}
\subsection{Basic experiments}
\label{sec:pre-experiment}
In the experimental environment described in Section \ref{sec:env}, I divided $D_{train}$ into 20 equal numbers (A:B:C:D:E = 20:20:20:20:20). In the SR mixed environment, mixed dataset $D_{train}^{(mixed)}$ is given as training data. Furthermore, 100 subjects' data $D_{train}^{6.25(unified)}$, which is generated from $D_{train}^{(mixed)}$ via downsampling, is given as a pseudo-reproduced sampling rate unified environment. I compare the $D_{test}^{6.25}$ estimation accuracy between the model trained by $D_{train}^{(mixed)}$ and $D_{train}^{6.25(unified)}$ in order to verify the effect of the SR mixed environment.

\subsubsection{The effect of SR mixed environment}
I compare not only a representation learning method using a deep learning model $M_{org}$ but also a machine learning method using HCF. In this study, HCF adopted 51 features that were used in a previous study \cite{hasegawa2019}. Each feature is standardized on the basis of the training data. The machine learning algorithms are SVM (Support Vector Machine) \update{\cite{Cortes1995}}, kNN (k-nearest neighbor) \update{\cite{Cover1967}}, RF (Random Forests) \update{\cite{Breiman2001}}, and DNN (Deep Neural Networks). The architecture of DNN, which is composed of fully-connected layer (2000 units) and a dropout layer stacked twice, is also the same as that in a previous study.

First, the results of comparing the algorithms using HCF are shown in Table \ref{table:hcfresult}. This table indicates the average estimation accuracy for all the sampling rates ($D_{test}^{100}$ to $D_{test}^{6.25}$). As a result, DNN achieved the best accuracy of 70.2\%. Therefore, I will refer to DNN as $M_{HCF-DNN}$ in the following discussion.
\begin{table}[b] 
\caption{Comparison of the estimation accuracy between machine learning algorithms using HCF (the average value of all sampling rates data of $D_{test}$).}
\label{table:hcfresult}
\scalebox{1.}{
\hbox to\hsize{\hfil
\begin{tabular}{c|c} \hline \hline
Algorithm &	Accuracy [\%] \\ \hline
SVM &	68.9 \\
kNN &	62.8 \\
RF &	69.4 \\
DNN &	\ub{70.2} \\ \hline
\end{tabular}\hfil}
}
\end{table}
\begin{figure}[b]
\begin{center}
\includegraphics[width=8.5cm]{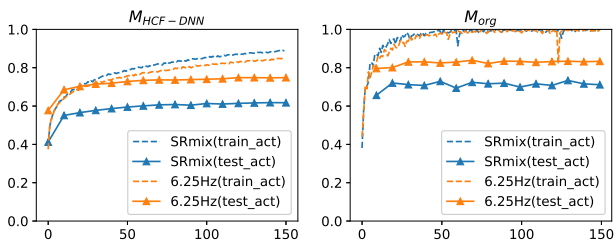}
\caption{Changes in estimation accuracy during training and testing of each model. ``SRmix'' indicates the estimation accuracy using each model trained by $D_{train}^{(mixed)}$, and ``6.25Hz'' indicates that one trained by $D_{train}^{6.25(unified)}$.}
\label{fig:pre_result}
\end{center}
\end{figure}

Next, the results of comparing the estimation accuracy of $D_{test}^{6.25}$ in the mixed SR environment ($D_{train}^{(mixed)}$) and the environment in which SR is unified to 6.25 Hz ($D_{train}^{6.25(unified)}$) are shown in Fig. \ref{fig:pre_result}. Focusing on test\_act, which is the activity-recognition accuracy of $D_{test}^{6.25}$, the accuracy is higher for both models when the training data with a unified sampling rate of 6.25 Hz is used. Therefore, it was confirmed that the SR mixed environment had a negative effect on the estimation accuracy both with the conventional model and with the deep learning model. In addition, it can be confirmed from the figure that $M_{org}$ is about 10\% more accurate than $M_{HCF-DNN}$.
\begin{figure}[b]
\begin{center}
\includegraphics[width=8.5cm]{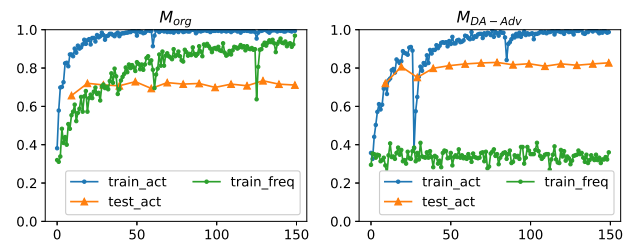}
\caption{Changes in estimation accuracy of activity recognition and sampling rate detection of each model. }
\label{fig:pre_frequency}
\end{center}
\end{figure}

\subsubsection{Discussion on estimation accuracy of sampling rate}
I compare the sampling rate estimation accuracy between the conventional method $M_{org}$ and my proposed method $M_{DA-Adv}$. Although the original $M_{org}$ model does not classify the sampling rate, using the $M_{DA-Adv}$ model (see Fig. \ref{fig:adversarial}) without conducting DA via down sampling and maximizing $L_D$ for $E$ enables us to verify the performance of the sampling rate estimation, because $E$ and $C$ in this model are equivalent to $M_{org}$.

Fig. \ref{fig:pre_frequency} shows the results of the sampling rate estimation. Focusing on the train\_freq of $M_{org}$, I find that the estimation accuracy of the sampling rate significantly increases as the training progresses. The best accuracy is 96.9\%; this result indicates that a unique feature representation for each sampling rate was extracted through the optimization using cross-entropy loss for the activity recognition. In contrast, when focusing on the $M_{DA-Adv}$, the estimation accuracy of the sampling rate is stable at about 30\%. This result is equivalent to the accuracy when all data is predicted to be at 6.25 Hz, which is the majority after conducting DA. Therefore, my proposed model could control the improvement of the estimation accuracy of the sampling rate, because an adversarial network for sampling rates is suitably implemented.

\subsubsection{Evaluation indicators of activity-recognition accuracy}
Focusing on the degree of convergence of the activity-recognition accuracy in Fig. \ref{fig:pre_frequency}, both models converge around 50 epochs, which indicates that the training of 150 epochs is sufficient. I therefore conducted the subsequent experiments using 150 epochs of training. Test accuracy is often discussed in terms of the accuracy of the final epoch, but, in this study, I will use the maximum accuracy of all the tests. This indicator does not arbitrarily distort the evaluation, because it does not waver and is basically stable, as shown in Fig. \ref{fig:pre_result}. 

On the other hand, if $M_{DA-Adv}$ is used, a sudden decrease in the training accuracy can occur at 30 and 80 epochs. When evaluated in the final epoch, the accuracy is greatly decreased when this phenomenon happens near the last epoch, which makes it difficult to create a uniform comparison. Therefore, I decided to use the maximum test accuracy while training for comparison in this study. Note that adopting this evaluation index did not overturn the considerations described in the following sections. I also confirm the estimation accuracy at the final epochs, which did not greatly differ.

Random sampling of the $D_{train}$ and $D_{test}$ subjects affects the test accuracy, depending on which subject is selected. Therefore, all subsequent analyses will be conducted as 10 trials for each pattern by changing the randomly sampled subjects, and I will then discuss the average accuracy thereof. 

\subsubsection{Discussion on estimation accuracy of human activity}
Table \ref{table:compare_acc} shows the experimental result the horizontal axis of which indicates the sampling rates of the test data $D_{test}$, and each value, which indicates the activity-recognition accuracy. ``Trained SR'' means that each sampling rate is included in training data $D_{train}^{(mixed)}$, and ``Unknown SR'' means that each sampling rate are not included in the training data, namely, un-trained sampling rate.
\begin{table*}[t] 
\caption{Average of estimation accuracy for each method using 100 persons as training data over 10 trials.}
\label{table:compare_acc}
\scalebox{1.}{
\hbox to\hsize{\hfil
\begin{tabular}{c|ccccc|ccccc} \hline \hline
 &	\multicolumn{10}{c}{Sampling rate of $D_{test}$} \\
 &	\multicolumn{5}{c|}{Trained SR} &	\multicolumn{5}{c}{Unknown SR} \\
model &	100 Hz &	50 Hz &	25 Hz &	12.5 Hz &	6.25 Hz &	33.3 Hz &	20 Hz &	10 Hz &	5 Hz &	4 Hz \\ \hline
$M_{HCF-DNN}$ &	70.5\% &	72.4\% &	73.9\% &	72.4\% &	62.2\% &	73.7\% &	73.8\% &	70.2\% &	57.1\% &	45.3\% \\
$M_{org}$ &	85.5\% &	85.8\% &	85.9\% &	84.8\% &	74.7\% &	86.0\% &	85.9\% &	83.9\% &	59.2\% &	39.2\% \\
$M_{DA-Adv}$ &	\ub{86.8\%} &	\ub{87.0\%} &	\ub{87.1\%} &	\ub{86.7\%} &	\ub{82.5\%} &	\ub{87.2\%} &	\ub{87.2\%} &	\ub{86.3\%} &	\ub{65.8\%} &	\ub{46.0\%} \\ \hline
\end{tabular}\hfil}
}
\end{table*} 

According to this table, my proposed method achieved the highest accuracy for all of the sampling rates of $D_{test}$. The difference in the accuracy between $M_{DA-Adv}$ and $M_{org}$ was 1.3\% (100 Hz), 1.2\% (50 Hz), 1.2\% (25 Hz), 1.9\% (12.5 Hz), and 7.9\% (6.25 Hz), respectively. Since DA via downsampling expanded the low-sampling-rate data, the accuracy for the low-sampling-rate data was especially improved. I also found that the trend of estimation accuracy for the ``Unknown SR'' data was same as for the ``Trained SR'' data.

\subsubsection{F-measure for each activity}
\begin{figure*}[t]
	\begin{tabular}{cc}
		\begin{minipage}[t]{0.5\hsize}
			\centering
			\includegraphics[keepaspectratio, scale=0.5]{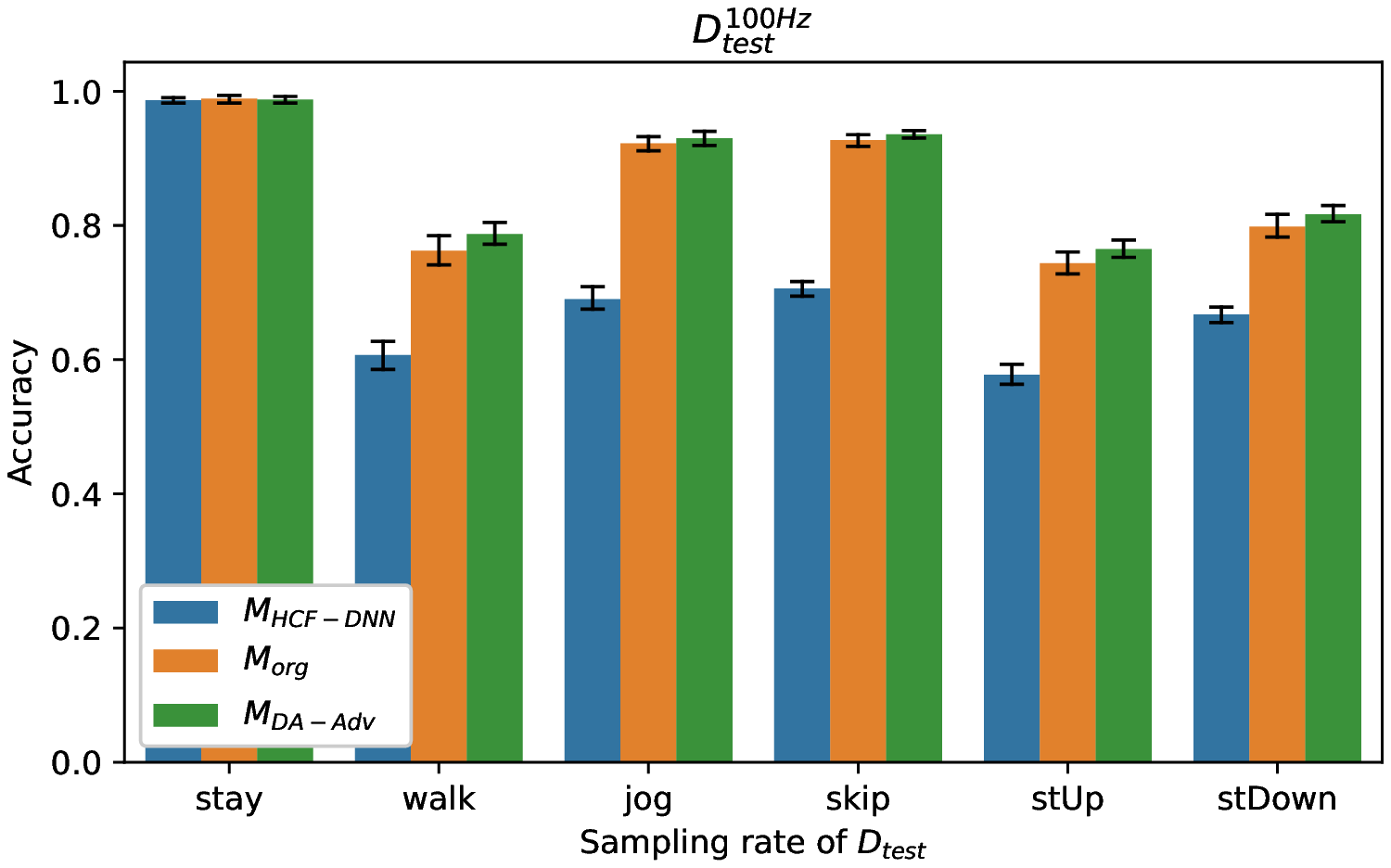}
		\end{minipage} &
		\begin{minipage}[t]{0.5\hsize}
			\centering
			\includegraphics[keepaspectratio, scale=0.5]{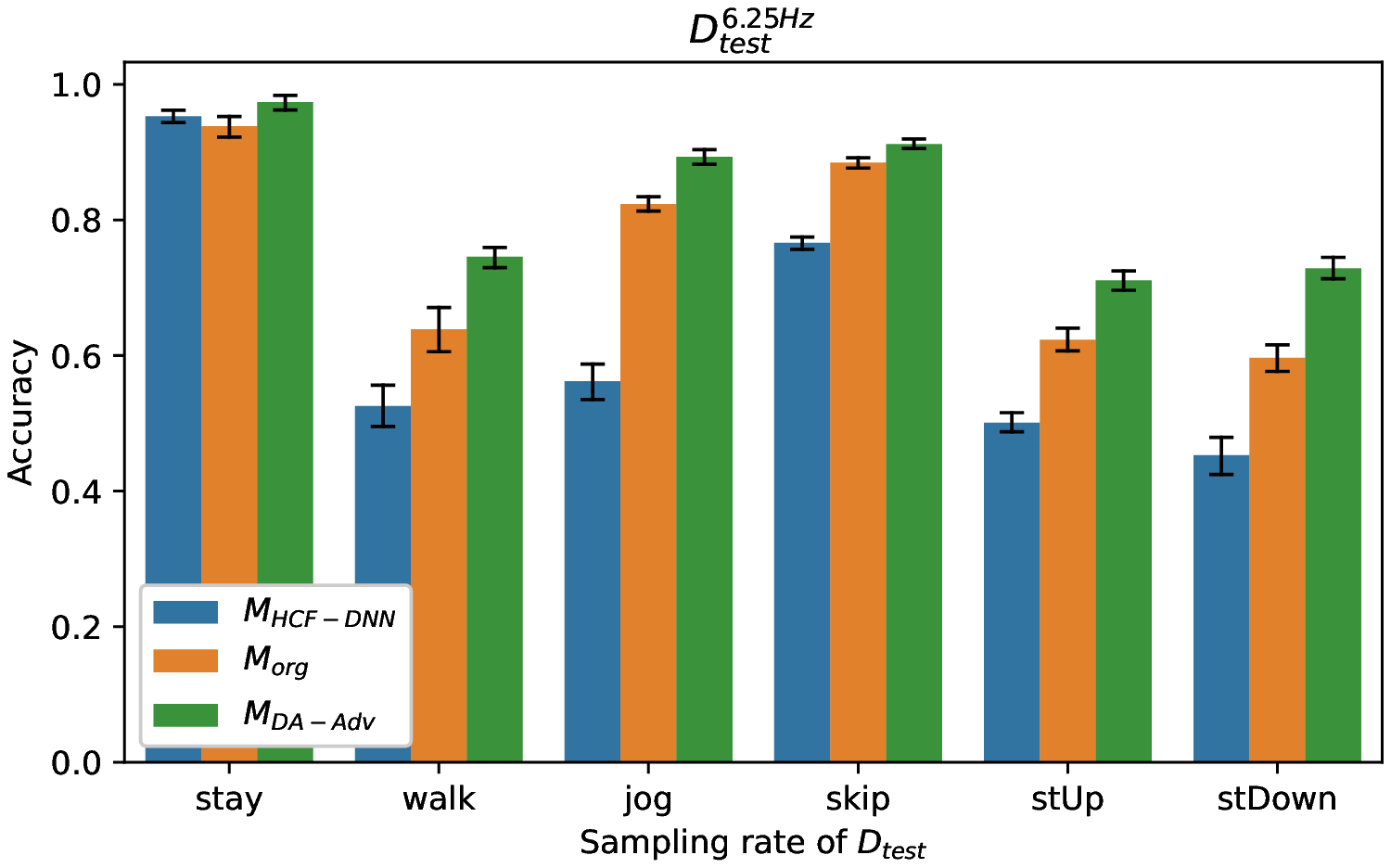}
		\end{minipage}
	\end{tabular}
		\caption{Average of f-measure for each activity using 100 persons as training data over 10 trials.}
		\label{fig:compare_fmeasure}
\end{figure*}

The above-mentioned results showed almost the same trends for 100 Hz and 12.5 Hz; therefore, I focus on only the results of the 100 Hz and 6.25 Hz data in the following experiments. Fig. \ref{fig:compare_fmeasure} shows the comparison result of the f-measures of each activity for the test data. Focusing on the difference between $M_{org}$ and $M_{DA-Adv}$, the trend is almost the same as in $D_{test}^{100Hz}$, but the proposed method is slightly more accurate for each activity. Focusing on $D_{test}^{6.25Hz}$, I can see that the general trend is similar, but the improvement in accuracy for walk, stUp, and stDown is particularly noticeable. In other words, the proposed method contributes to improving the activity-recognition accuracies the activity of which is particularly difficult to discriminate.

\begin{figure*}[t]
\begin{center}
\includegraphics[width=19cm]{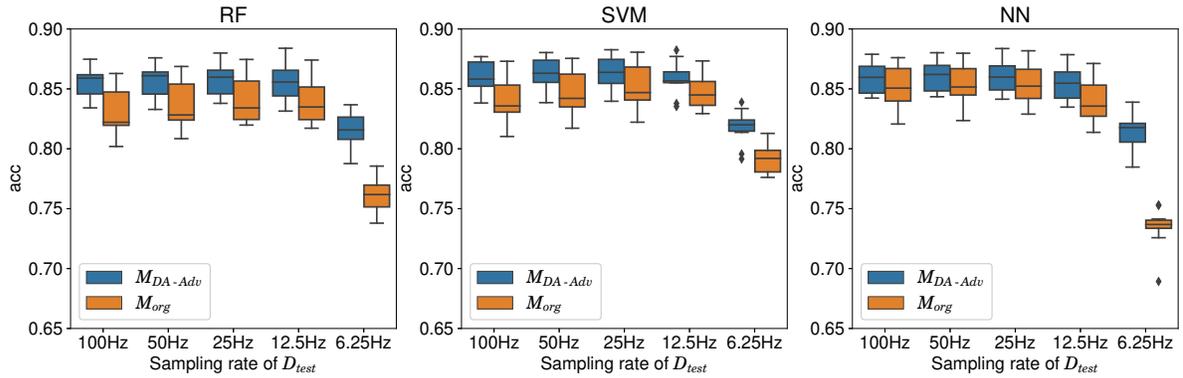}
\caption{Changes in estimation accuracy for each classifier $C$. }
\label{fig:machinelearning}
\end{center}
\end{figure*}

\update{
\subsubsection{Robustness against the difference of classifier}
To verify whether the encoder $E$ could acquire a feature representation $z$ that is robust to different sampling rates, I determined that my proposed method worked well when the classifier $C$ was a conventional machine learning method (SVM or RF). In other words, I first trained the model (Fig. \ref{fig:conventional} or \ref{fig:adversarial}), second calculated feature map $z$, namely, $E(D_{train})$, third trained a conventional machine learning model $C$ using $E(D_{train})$, and finally evaluated the estimation accuracy of $C$ using $E(D_{test})$. From the experimental results shown in Fig. \ref{fig:machinelearning}, we found that the feature representation extracted by my proposed method, $M_{DA-Adv}$, was more accurate then the conventional method, $M_{org}$, in activity recognition under all sampling rates, regardless of the kind of classifier. Note that the results of neural network (NN) were less than in Table \ref{table:compare_acc} in spite of using the same method because Fig. \ref{fig:machinelearning} showed the accuracy of the final epoch in the training and Table \ref{table:compare_acc} showed the accuracy of the best epoch in the training.
}

\subsection{Ablation study}
My proposed method combines DA by downsampling with an adversarial network. As an ablation study, I implemented an individual function in addition to the $M_{org}$ and $M_{DA-Adv}$. $M_{DA-org}$ is the method combining DA with the conventional model $M_{org}$, and $M_{Adv}$ is the simple adversarial network without DA. 

The results of the ablation study are shown in Table \ref{table:ablation_result}, where each method was treated independently. The results showed that the estimation accuracy was improved by DA, and the combination of the adversarial network further improved the accuracy for both $D_{test}^{100}$ and $D_{test}^{6.25}$.
\begin{table}[t] 
\caption{Ablation study on accuracy of activity recognition of my proposed method.}
\label{table:ablation_result}
\scalebox{1.}{
\hbox to\hsize{\hfil
\begin{tabular}{c|cc|cc} \hline \hline
\multirow{2}{*}{Model} &	\multirow{2}{*}{DA} &	Adversarial &	\multirow{2}{*}{$D_{test}^{100}$ [\%]} &	\multirow{2}{*}{$D_{test}^{6.25}$ [\%]} \\ 
 &	 &	network &	 &	\\ \hline
$M_{org}$ &	- &	- &	85.5($\pm$1.6) &	74.7($\pm$1.7) \\
$M_{DA-org}$ &	o &	- &	86.1($\pm$1.5) &	82.3($\pm$1.4) \\
$M_{Adv}$ &	- &	o &	85.1($\pm$2.5) &	74.8($\pm$2.1) \\
$M_{DA-Adv}$ &	o &	o &	\ub{86.8($\pm$1.4)} &	\ub{82.5($\pm$1.3)} \\ \hline
\end{tabular}\hfil}
}
\end{table}

\subsubsection{Effects of the number of subjects in the training data}
Based on the results of the ablation study, I discuss under what conditions each function works effectively. In previous experiments, the number of subjects under each sampling rate in $D_{train}$ was 20, for a total of 100 subjects. In this section, to discuss the effect of reducing the number of subjects in $D_{train}$ from 100, I evaluated the cases in which the number of subjects in $D_{train}$ was 100 (A:B:C:D:E = 20:20:20:20:20), 50 (10:10:10:10:10), and 25 (5:5:5:5:5). 

\begin{figure*}[t]
	\begin{tabular}{cc}
		\begin{minipage}[t]{0.5\hsize}
			\centering
			\includegraphics[keepaspectratio, scale=0.5]{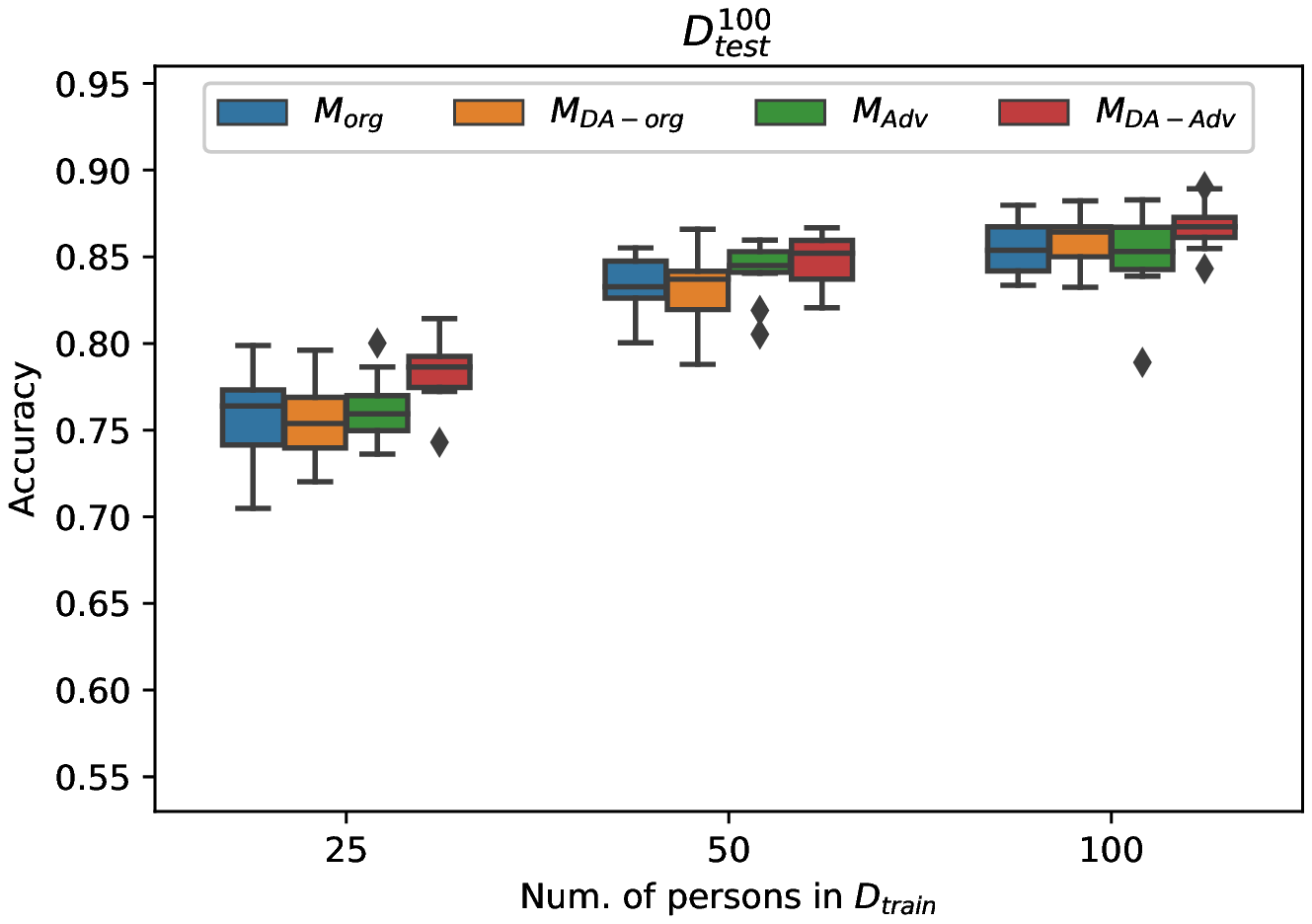}
		\end{minipage} &
		\begin{minipage}[t]{0.5\hsize}
			\centering
			\includegraphics[keepaspectratio, scale=0.5]{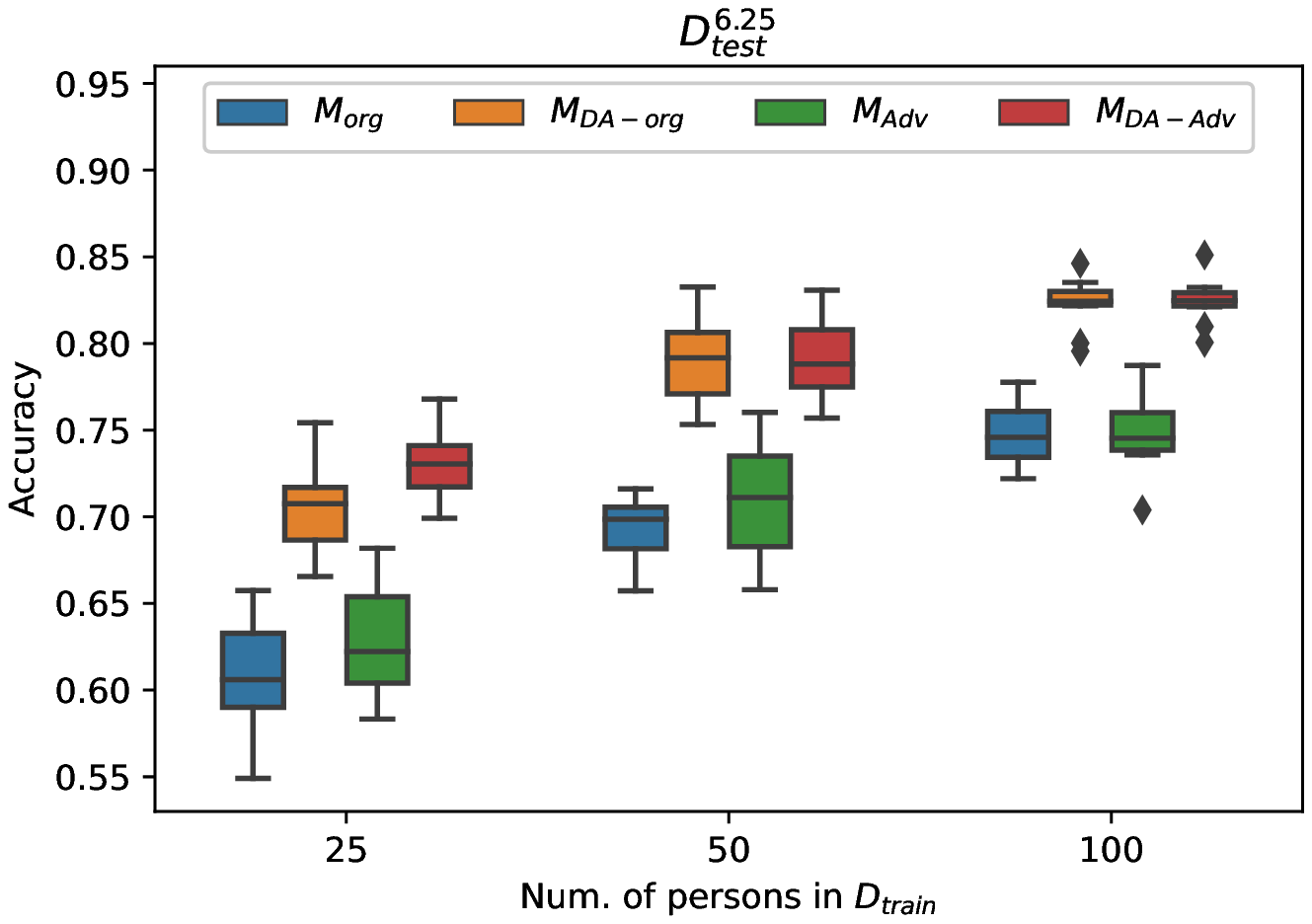}
		\end{minipage}
	\end{tabular}
	\caption{Comparison of activity-recognition accuracy for each method when the number of subjects in training data is changed.}
	\label{fig:compare_nop}
\end{figure*}

Fig. \ref{fig:compare_nop} shows the box plots, which summarize the accuracy of the 10 trials. The horizontal axis indicates the number of subjects in the training data, and the vertical axis indicates the activity-recognition accuracy. Focusing on the left $D_{test}^{100}$, my proposed method achieved the highest accuracy in the three cases. Furthermore, we found that the degree of improvement from my method increased with a decreasing number of subjects in the training data. Focusing on the right $D_{test}^{6.25}$, the accuracy of $M_{DA-org}$ was almost the same as that of $M_{DA-Adv}$ when the number of subjects was 50 or 100. In the case where the training data was sufficiently collected, the model could acquire sufficient feature representation because my DA method generated 6.25 Hz data for all the subjects. On the other hand, when the number of subjects in the training data is 25, the accuracy is improved by about 2.5\% by combining DA with the adversarial network. Therefore, the sampling rate adversarial network can improve the estimation accuracy when the number of subjects in the training data is low.

\subsubsection{Effects of the bias of the number of subjects in the training data}
In the above-mentioned experiments, I assumed that the sensor data for each sampling rate was collected at the same ratio. In this section, I discuss the effects of the difference in the ratio of the number of subjects for each sampling rate in the training data. I evaluated three scenarios: \um{Low}, where more training data at a lower sampling rate is available (A:B:C:D:E = 5:5:5:5:30), \um{Even}, where the same number of subjects for training data at each sampling rate is available (10:10:10:10:10), and \um{High}, where more training data at a higher sampling rate is available (30:5:5:5:5).

\begin{figure*}[t]
	\begin{tabular}{cc}
		\begin{minipage}[t]{0.5\hsize}
			\centering
			\includegraphics[keepaspectratio, scale=0.5]{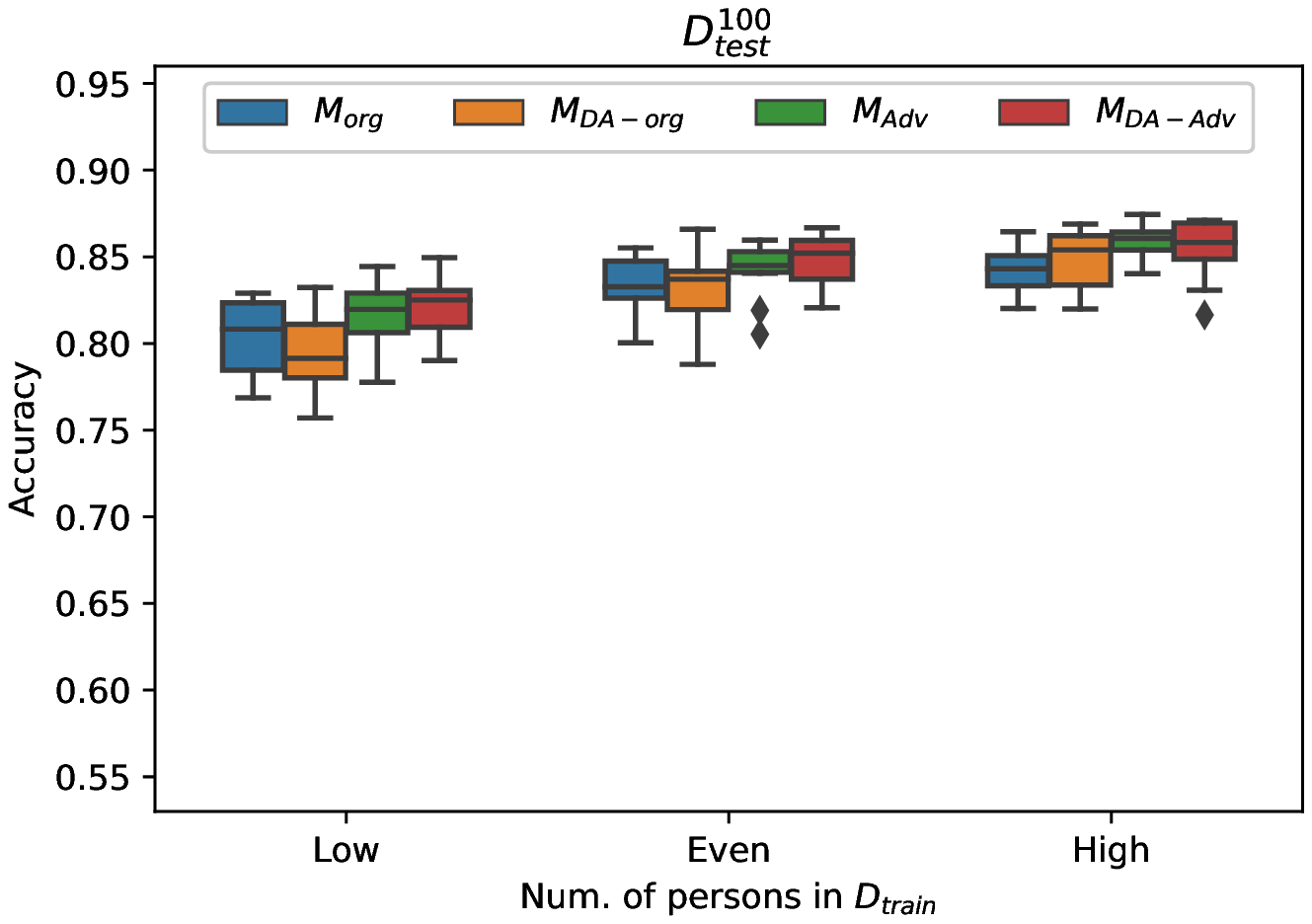}
		\end{minipage} &
		\begin{minipage}[t]{0.5\hsize}
			\centering
			\includegraphics[keepaspectratio, scale=0.5]{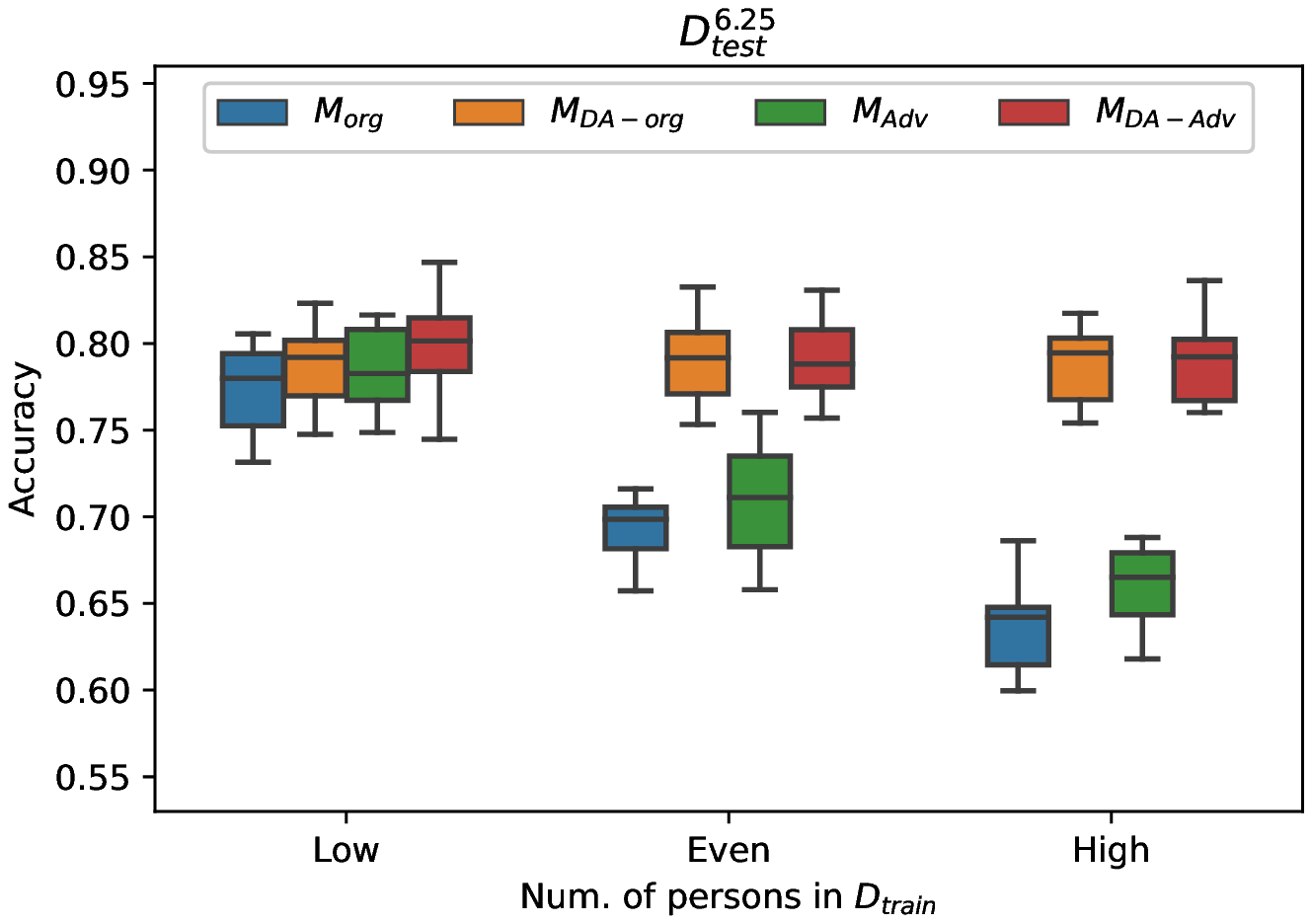}
		\end{minipage}
	\end{tabular}
	\caption{Comparison of activity-recognition accuracy for each method when the bias of the number of subjects in training data is changed.}
	\label{fig:compare_ratio}
\end{figure*}

Fig. \ref{fig:compare_ratio} shows the box plots, which summarize the accuracy of the 10 trials. The horizontal axis indicates each scenario, and the vertical axis indicates the activity-recognition accuracy. Focusing on the left $D_{test}^{100}$, the adversarial networks ($M_{Adv}$ and $M_{DA-Adv}$) achieved better accuracy than the other models. Since DA by downsampling could not expand the 100 Hz training data, DA did not improve the accuracy, while the adversarial network did. Focusing on the right $D_{test}^{6.25}$, similar to the previous section, DA remarkably improved the estimation accuracy. Furthermore, in the \um{Low} scenario, combining DA with the adversarial network improved the accuracy. A reason for this is that the effect of DA was lower because the \um{Low} scenario includes a relatively limited amount of high-sampling rate data.

\subsubsection{Estimation accuracy of sensor data actually measured at unknown sampling rate}
Since this data is pseudo-reproduction based on downsampling, the downsampled data can be different from the actual measured sensor data. In addition to the data measured at a 100 Hz sampling rate, the HASC dataset contains sensor data that is measured at a 40 Hz sampling rate. I verify the estimation accuracy using 20 subjects' $D_{org}^{40}$, which is not a pseudo-reproduction by downsampling.
\begin{figure*}[t]
	\begin{tabular}{cc}
		\begin{minipage}[t]{0.5\hsize}
			\centering
			\includegraphics[keepaspectratio, scale=0.5]{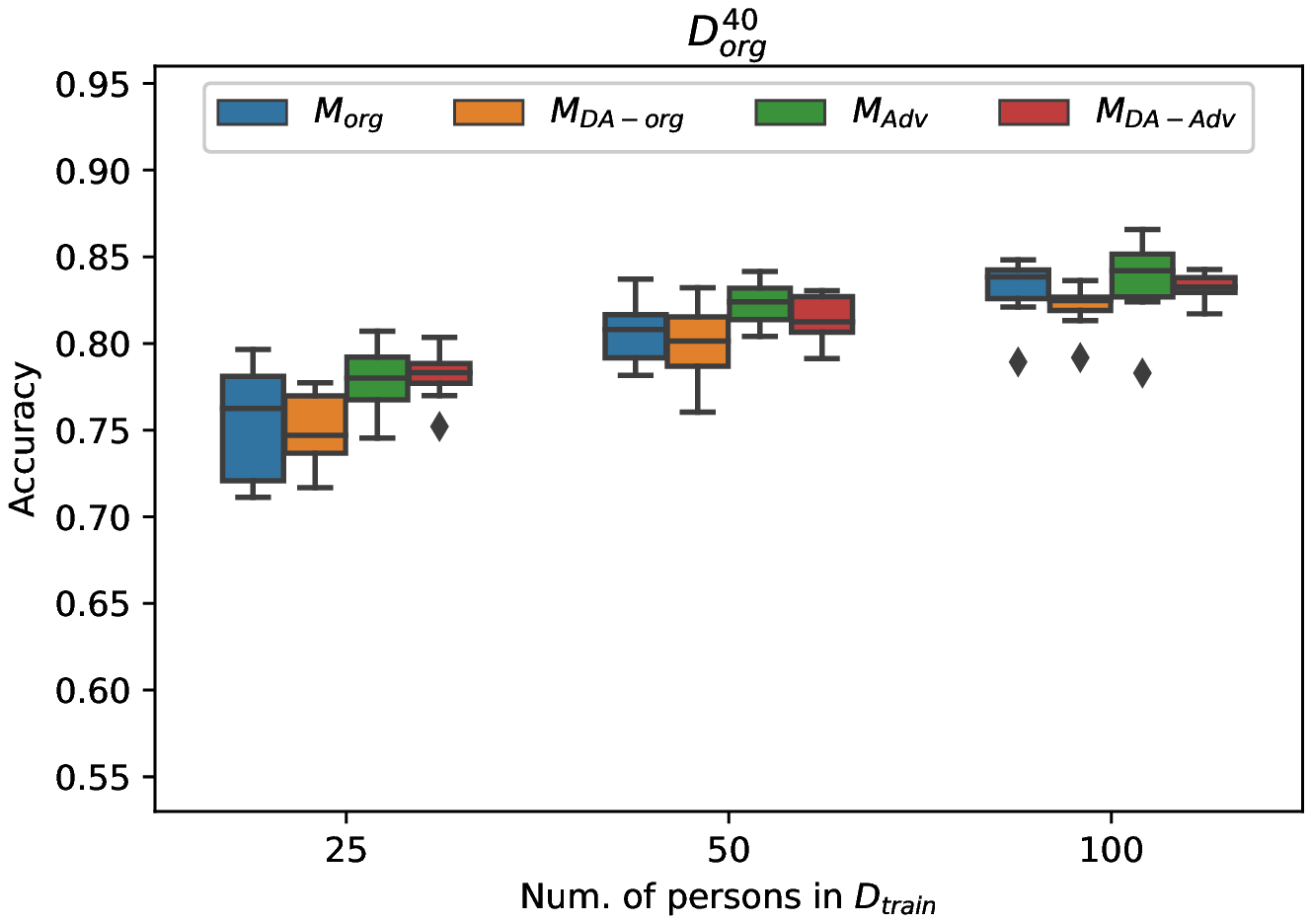}
		\end{minipage} &
		\begin{minipage}[t]{0.5\hsize}
			\centering
			\includegraphics[keepaspectratio, scale=0.5]{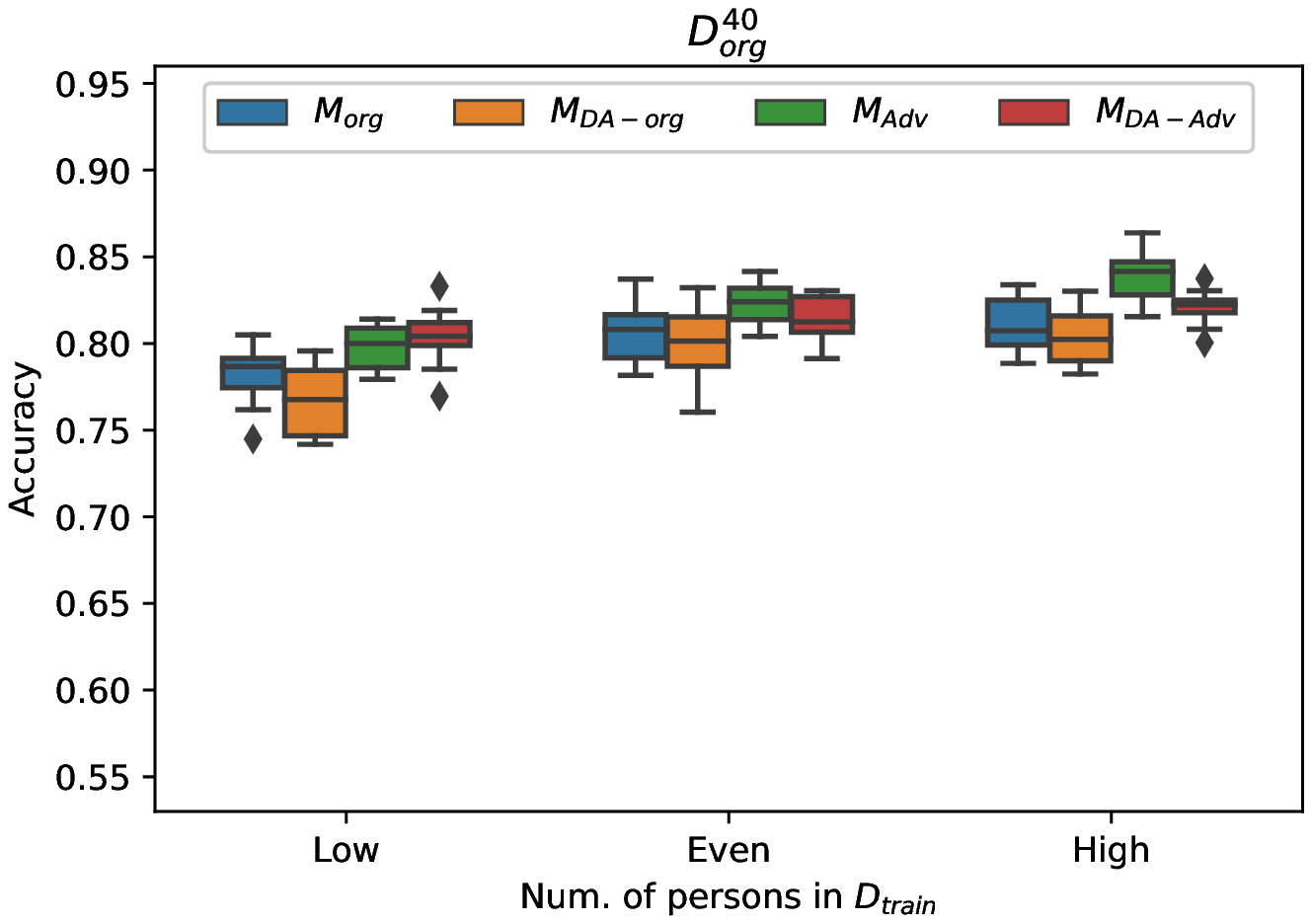}
		\end{minipage}
	\end{tabular}
	\caption{Comparison of estimation accuracy when the test data measured at unknown sampling rate in actual environment}
	\label{fig:compare_org}
\end{figure*}

According to Fig. \ref{fig:compare_org} which shows the experimental results, $M_{Adv}$ showed the best performance for $D_{org}^{40}$, which was not a pseudo-reproduction, even though data augmentation by downsampling is sometimes effective, such as when the number of subjects in the training data is low and more training data at a lower sampling rate is available (\um{Low} in right portion of Fig. \ref{fig:compare_org}). This may be because the sampling rate of the test data is relatively high (40 Hz), and the pseudo-data augmentation by downsampling may adversely affect the estimation accuracy. In this study, downsampling was performed by an LPF and thinning. Because this method could not correctly reproduce the actual sensor data measured at low sampling rates, I will try to change the downsampling method as a future work. By using a downsampling method adapted to the actual environment, the accuracy of my proposed method can be further improved. Furthermore, in the previous section, we found that DA particularly contributed to improving the estimation accuracy for the test data measured at a low sampling rate ($D_{test}^{6.25}$). Because I could not validate the effectiveness of DA for very low sampling rate data (i.e. 6.25 Hz), which was not included in the dataset this time, I will try to investigate the effectiveness of DA for very low sampling data measured in the real environment as a future work.

\subsection{Limitations and usefulness}
The following two points are the limitations of this study.
\begin{itemize}
\item \textbf{Activity recognition using a smartphone.} I conducted the experiments using the HASC corpus \cite{HASC}, which is comprised of sensor data measured by smartphones for basic activity recognition. Therefore, the evaluation of the effectiveness of my proposed method is thus far limited to the range of activity recognition using smartphones. According to the conventional method $M_{org}$ in Table \ref{table:compare_acc}, the estimation accuracy greatly decreased between 12.5--6.25 Hz. In basic activity recognition, such as HASC, this result indicates the importance of the features of the low-sampling-rate component; namely, a 12.5 Hz sampling rate is sufficient to recognize the basic activity. Therefore, I will verify the effectiveness of my proposed method for activity-recognition tasks in which high-frequency components are particularly important for estimation.
\item \textbf{Pseudo-reproduction by downsampling.} In this study, I pseudo-reproduced the dataset for each sampling rate via downsampling using an LPF and thinning. Since this may be different from real-world measurements, it was suggested that this may lead to a decrease in the estimation accuracy for $D_{org}^{40}$. I verified the effectiveness of my proposed method, the sampling-rate adversarial network; however, it is necessary to consider a downsampling method that is adapted to the actual environment.
\end{itemize}

I discuss the usefulness of this study based on its limitations. The effectiveness of my proposed method was demonstrated using the HASC dataset, but the data measured at approximately a 12.5 Hz sampling rate was sufficient for basic activity recognition. Since the majority of HASC datasets are measured at 100 Hz, the usefulness of this study is questionable. However, considering the actual situation of activity recognition, continuously measuring high-sampling-rate sensor data, such as at 100 Hz, is not feasible due to battery and CPU consumption. My proposed method will therefore be useful when using the activity recognition system in the future. In addition, my experimental results showed that my proposed method could improve estimation accuracy using sensor data from many users---even if said data was measured in an SR mixed environment---rather than collecting data from a small number of subjects at a uniform sampling rate. My proposed method will help improving the accuracy of future research on activity recognition.

\section{Conclusion}
In this study, I focused on basic activity recognition via smartphone sensing in the context of a mixed sampling rate, and I developed an environmentally robust activity-recognition method. I proposed a sampling-rate adversarial network and data augmentation by downsampling. The adversarial model has two components: activity recognition and discrimination of the sampling rate. Adversarial training of both of these components makes it possible to acquire a feature representation that is robust to different sampling rates. The effectiveness of the proposed method was assessed with the HASC dataset of basic-activity recognition using smartphones. As a result, I clarified that the method combining the sampling-rate adversarial network and data augmentation by downsampling works better than conventional methods. I only evaluated my proposed method using the HASC dataset that included sensor data measured at various sampling rates, since there are few datasets that include various sampling rate data. I will construct other datasets that include sensor data measured at various sampling rates for evaluation, and verify my proposed method for other sensing tasks as future work.

\begin{IEEEbiography}[{\includegraphics[width=1in,height=1.25in,clip,keepaspectratio]{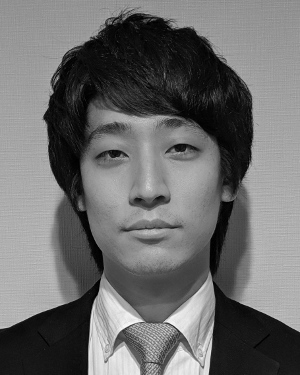}}]{Tatsuhito Hasegawa} (M'13) received Ph.D. degree in engineering from Kanazawa University, Ishikawa, in 2015. From 2011 to 2013, he was a system engineer with the Fujitsu Hokuriku Systems Limited. From 2014 to 2017, he was an Assistant with the Tokyo Healthcare University. Since 2017, he has been a Senior Lecturer with the Graduate School of Engineering, University of Fukui. His research interests include human activity recognition, applying deep learning, and intelligent learning support system. He is a member of IPSJ, and JASAG.
\end{IEEEbiography}

\end{document}